\documentstyle[preprint,aps]{revtex}
\begin{document}
\baselineskip=17pt
\title{Quantum Coherence and W$_{\infty }\times $SU(2) Algebra \\
in Bilayer Quantum Hall Systems }
\author{Z. F. Ezawa}
\address{ Department of Physics, Tohoku University, Sendai 980,
Japan } 
\maketitle
\begin{abstract} We analyze the bilayer quantum Hall
(QH) system by mapping it to the monolayer QH system with spin
degrees of freedom. By this mapping the tunneling interaction
term is identified with the Zeeman term. We clarify the mechanism
of a spontaneous development of quantum coherence based on the
Chern-Simons gauge theory with the lowest-Landau-Level projection
taken into account. The symmetry group is found to be W$_{\infty
}\times $SU(2), which says that the spin rotation affects the
total electron density nearby. Using it extensively we construct
the Landau-Ginzburg theory of the coherent mode. Skyrmion
excitations are topological solitons in this coherent mode. We
point out that they are detectable by measuring the Hall current
distribution. 
y\newline {\sl PACS:}\ 73.40.Hm, 73.20.Dx, 73.40.-c, 75.10.-b
\newline 
\end{abstract}

\section{Introduction:}

In the 2-dimensional space entirely new phenomena can occur due
to its intrinsic topological structure. For instance, an electron
may be transmuted into a boson by making a charge-flux composite
in external magnetic field, which is called {\it composite
boson}. As a result electrons may condense without making Cooper
pairs. The fractional quantum Hall (QH) state is such a
condensate of composite bosons \cite{LGCSx}. When the spin
degrees of freedom are taken into account, a quantum coherence
develops spontaneously and turns the QH system into a {\it
quantum Hall ferromagnet}. Skyrmions \cite{Skyrmion,SondhiKaKi}
are new topological solitons in this mode, which have been
observed experimentally \cite{SkyrmExper}. In a certain bilayer
QH system an interlayer coherence develops spontaneously
\cite{EIjos} and Josephson-like phenomena are expected to occur
\cite{EIjos,EzaFerro}. Some characteristic properties of the mode
have already been observed experimentally\cite{Sheena}. The aim
of this paper is to clarify the mechanism of the spontaneous
development of quantum coherence in these two systems in a
unified manner.

To make a consistent theory of the fractional QH effect it is
necessary to make the lowest-Landau-level (LLL) projection
\cite{refLLL}, which has so far been used only within the
single-mode approximation (SMA) \cite{refLLL,MG}. In this paper,
proposing a bosonic Chern-Simons (CS) gauge theory with the LLL
projection, we apply it to the study of two-component QH systems.
We map explicitly the bilayer system to the monolayer system with
spin degrees of freedom. This allows us to analyze both systems
in a unified way. This also helps us to understand the proper
roles of the capacitance and tunneling terms in the bilayer
system. In particular, the tunneling interaction term corresponds
to the Zeeman term.

It is one of our main results that the U(1) symmetry is not
broken spontaneously in spite of bose condensation. On the other
hand, the SU(2) symmetry is found to be broken spontaneously,
yielding a coherent mode. After the LLL projection the dynamics
is governed by the W$_{\infty }\times $SU(2) algebra, which says
that the spin rotation affects the total electron density.
Namely, when we make a spin rotation the total electron density
nearby is modulated. Using this property, we derive the
Landau-Ginzburg (LG) theory of the coherent mode. Skyrmions
\cite{Skyrmion} are collective excitations in this mode
\cite{SondhiKaKi}. We show that they are detectable by measuring
the Hall current distribution. We also present a systematic
method to calculate the current and static correlation functions.
Our field-theoretical analysis confirms some of the results made
in the SMA \cite{MG}. In this paper we use the natural unit
$c=\hbar =1$ and take the length unit to be the magnetic radius
$\ell_{B}=\sqrt {\hbar c/eB}$ with $B$ the magnetic field. The
Landau-level energy gap is given by $\hbar \omega _{c}=1/M$.
Hence, the LLL projection is associated with limit $M\rightarrow
0$. 

\section{SU(2) Spin Structure}

We consider a bilayer electron system in strong magnetic field.
We denote the electron field at the layer $\alpha (=1,2)$ by
$\psi _{\alpha }$. The Hamiltonian is
\begin{equation} H = {1\over 2M}\sum _{\alpha }\int d^{2}x \psi
^{\dagger }_{\alpha }(\bbox{x})(P_{x}^{2}+P_{y}^{2})\psi _{\alpha
}(\bbox{x}) + {1\over 2}\sum _{\alpha ,\beta }\int \!d^{2}xd^{2}y
V_{\alpha \beta }(\bbox{x}-\bbox{y})\psi ^{\dagger }_{\alpha
}\psi _{\alpha }(\bbox{x})\psi ^{\dagger }_{\beta }\psi _{\beta
}(\bbox{y}), \label{HamilBL} 
\end{equation}
where $P_{k}$ is the covariant derivative, $P_{k}=-i\partial
_{k}+eA_{k}^{\rm{ext}}$ with $A_{k}^{\rm{ext}}={1\over
2}\varepsilon _{kj}x_{j}B$, and $V_{\alpha \beta
}(\bbox{x}-\bbox{y})$ the Coulomb interaction. It is convenient
to introduce the symmetric field ($\psi ^{\uparrow }$) and
antisymmetric field ($\psi ^{\downarrow }$) by
\begin{equation}
\psi ^{\uparrow } = {1\over {\sqrt {2}}}(\psi _{1}+\psi _{2})
,\qquad \psi ^{\downarrow } = {1\over {\sqrt {2}}}(\psi _{1}-\psi
_{2}) , \label{LayerToSAS}
\end{equation}
and to construct a two-component field $\Psi =(\psi ^{\uparrow
},\psi ^{\downarrow })$. We may rewrite the Hamiltonian
(\ref{HamilBL}) as $H=H_{K}+H_{C}^{+}+H_{C}^{-}$ with
\begin{eqnarray} H_{K} &=&{1\over 2M}\int d^{2}x \Psi ^{\dagger
}(\bbox{x})(P_{x}-iP_{y})(P_{x}+iP_{y})\Psi (\bbox{x}) + {N\over
2}\hbar \omega _{c} , \label{SpinKinem}\\ H_{C}^{+}&=&{1\over
2}\int d^{2}xd^{2}y V_{+}(\bbox{x}-\bbox{y})\rho (\bbox{x})\rho
(\bbox{y}), \label{SpinCoulo}\\ H_{C}^{-}&=&{1\over 2}\int
d^{2}xd^{2}y
V_{-}(\bbox{x}-\bbox{y})S^{1}(\bbox{x})S^{1}(\bbox{y}),
\label{SpinCouloCC} 
\end{eqnarray}
where $V_{\pm }=V_{11}\pm V_{12}$ and $N$ is the total electron
number in the system. We have introduced the SU(2) generator
$S^{a}(\bbox{x})={1\over 2}\Psi ^{\dagger }\tau ^{a}\Psi $ with
$\tau ^{a}$ the Pauli matrices. The total density is $\rho
(\bbox{x})=\Psi ^{\dagger }\Psi =\rho _{1}+\rho _{2}=\rho
^{\uparrow }+\rho ^{\downarrow }$. Note that
$2S^{1}(\bbox{x})=\rho _{1}-\rho _{2}$ and $2S^{3}(\bbox{x})=\rho
^{\uparrow }-\rho ^{\downarrow }$. The antisymmetric Coulomb term
$H_{C}^{-}$ describes the capacitance energy stored between the
two layers; see (\ref{EnergChang}).

The tunneling interaction is described by
\begin{equation} H_{Z} = -{1\over 2}\Delta_{\rm{SAS}}\int d^{2}x
(\psi _{1}^{\dagger }\psi _{2} + \psi _{2}^{\dagger }\psi _{1}) =
-{1\over 2}\Delta_{\rm{SAS}} \int d^{2}x (\psi ^{\uparrow \dagger
}\psi ^{\uparrow }- \psi ^{\downarrow \dagger }\psi ^{\downarrow
}), 
\end{equation}
with $\Delta_{\rm{SAS}}$ the symmetric-antisymmetric energy gap.
We may rewrite this as
\begin{equation} H_{Z} = -\lambda \int d^{2}x S^{3}(\bbox{x}),
\label{SpinZeema}
\end{equation}
where we have set $\lambda \equiv \Delta_{\rm{SAS}}$. We start
with the regime where the Coulomb term $H_{C}^{+}$ dominates the
dynamics so that we can treat the capacitance term $H_{C}^{-}$
and the tunneling term $H_{Z}$ as a perturbation. In this
case the Halperin $(m,m,m)$ phase \cite{HalperinC} is realized
and an interlayer coherence develops spontaneously \cite{EIjos}.
The capacitance term is made small when the interlayer distance
$d$ is made small compared with the magnetic length $\ell _{B}$:
In particular, $H_{C}^{-}\rightarrow 0$ as $d\rightarrow 0$ and
$H_{C}^{-}\rightarrow H_{C}^{+}$ as $d\rightarrow \infty $. The
unperturbed system has the symmetry group SU(2).

We now consider the monolayer system with spin degrees of
freedom. The electron is described by a two-component field $\Psi
=(\psi ^{\uparrow },\psi ^{\downarrow })$ with up and down spins.
It is obvious that the Hamiltonian is given by
$H=H_{K}+H_{C}^{+}+H_{Z}$ with the kinetic term
(\ref{SpinKinem}), the Coulomb term (\ref{SpinCoulo}) and the
Zeeman term (\ref{SpinZeema}): Here, $\lambda =g\mu _{B}B$ with
$g$ is the gyromagnetic factor and $\mu _{B}$ the Bohr magneton.
We also start with the regime where the Coulomb term $H_{C}^{+}$
dominates the dynamics so that we can include the Zeeman term
$H_{Z}$ as a perturbation. In this way the bilayer system is
mapped to the monolayer system with spins except for the
capacitance term $H_{C}^{-}$.

The spin operator generates a local SU(2) transformation,
$e^{-i{\cal O}}$, with
\begin{equation} {\cal O} = \sum _{a=1}^{3}\int d^{2}x
f^{a}(\bbox{x}) S^{a}(\bbox{x}) , \label{SpinGenerU}
\end{equation}
where $f^{a}(\bbox{x})$ is a real function. It acts on the SU(2)
field as
\begin{equation}
\Psi (\bbox{x})\quad\rightarrow \quad e^{-i{\cal O}}\Psi
(\bbox{x})e^{i{\cal O}} = \exp\bigl[i\sum f^{a}(\bbox{x}){\tau
^{a}\over 2}\bigr]\Psi (\bbox{x}) . \label{SpinRotat}
\end{equation}
It generates a spin texture on the ground state $|g\rangle $,
$|\Phi \rangle =e^{i{\cal O}}|g\rangle $. The spin texture is an
excited state since the system does not possess the local SU(2)
symmetry. 

\section{Composite Bosons}

We analyze the unperturbed system consisting of the kinetic
term $H_{K}$ and the Coulomb term $H_{C}^{+}$. To show a
spontaneous development of quantum coherence it is most
convenient to use the composite bosons. The composite boson field
$\phi _{\alpha }$ is defined \cite{EIjos} by an operator phase
transformation with a common phase $\Theta $,
$
\phi _{\alpha }(\bbox{x}) = e^{i\Theta (\bbox{x})}\psi _{\alpha
}(\bbox{x}) $.
By this transformation the covariant derivative is modified as,
\begin{equation} P_{k} \rightarrow \check{P}_{k} \equiv
-i\partial _{k}+eA_{k}^{\rm{ext}} + C_{k}, \label{MomenCS}
\end{equation}
where the field $C_{k}(\bbox{x})\equiv \partial _{k}\Theta
(\bbox{x})$ is the CS gauge field to be determined by the CS
constraint,
\begin{equation}
\varepsilon _{jk}\partial _{j}C_{k} = 2\pi m\rho ,
\label{CSconstSPN}
\end{equation}
in terms of the total density $\rho $. Here, $m$ is an odd
integer which makes the field $\phi _{\alpha }$ bosonic. The spin
operators and the total density are the same, $S^{a}={1\over
2}\Phi ^{\dagger }\tau ^{a}\Phi $ and $\rho =\Phi ^{\dagger }\Phi
$ where $\Phi =(\phi ^{\uparrow },\phi ^{\downarrow })$ in terms
of the symmetric and antisymmetric fields, as those in the
original electron theory.

From the Hamiltonian with the covariant derivative
$\check{P}_{k}$ the mean-field ground state is found to be
\begin{equation}
\langle \phi ^{\alpha }\rangle = e^{\varphi _{0}^{\alpha }}\sqrt
{\rho _{0}^{\alpha }} , \qquad C_{k}+eA_{k}^{\rm{ext}}=0 ,\qquad
(\alpha =\uparrow \downarrow ) \label{CSconstSpin} 
\end{equation}
with arbitrary constants $\varphi _{0}^{\alpha }$ and $\rho
_{0}^{\alpha }$ subject to $\rho _{0}^{\uparrow }+\rho
_{0}^{\downarrow }=2\rho _{0}$; here $2\rho _{0}$ stands for the
homogeneous background charge. Substituting it into the CS
constraint (\ref{CSconstSPN}) we find this ground state to realize
only at the filling factor $\nu \equiv {4\pi \rho _{0}\hbar
c/eB}={1/m}$. We have defined the Landau-level filling factor so
that it yields $\nu =1$ when all the up-spin electron sites are
filled.

The essential point is that there are many ground states
(\ref{CSconstSpin}) degenerate one another even at $\nu =1$: They
are indexed by $(\varphi _{0}^{\alpha }, \rho _{0}^{\alpha })$.
It is necessary to choose one of them as the ground state
$|g_{0}\rangle $ upon which to build the Hilbert space. This
breaks the SU(2) symmetry spontaneously and develops a quantum
coherence. Equivalently, the direction of the spin polarization
$S^{a}$ is spontaneously chosen, $\rho _{0}s^{a}_{0} \equiv
\langle g|S^{a}(\bbox{x})|g\rangle = \hbox{constant}$, where
$\sum _{a=1}^{3}(s_{0}^{a})^{2} = 1$. 
This is why the system is called a quantum Hall 
ferromagnet.  It is important that 
various densities are observable on the 
state $|g\rangle $, $\langle g|\rho _{1(2)}|g\rangle 
=\rho _{0}(1\pm s^{1}_{0})$ and $\langle g|
\rho ^{\uparrow(\downarrow)}|g\rangle =(1\pm s^{3}_{0})$.  In the 
monolayer system the Zeeman term fixes the polarization to be
\begin{equation} s^{a}_{0} \equiv {1\over \rho _{0}}\langle
g_{0}|S^{a}(\bbox{x})|g_{0}\rangle = \delta ^{a3},
\label{QHvacuu} 
\end{equation}
however small it may be $(\lambda \approx 0)$. In the bilayer
system the capacitance term is minimized by the choice of
$s^{1}=0$ and the tunneling term is minimized by $s^{3}=1$:
Hence, the resulting polarization is again given by
(\ref{QHvacuu}). This is the unique ground state of the total
system, where all electrons are in the symmetric state ($\psi
^{\uparrow }$). Because of this reason it remains to be the
ground state even if the Zeeman term (capacitance and tunneling
terms) is made large provided that it is not too large.

\section{Coherent State}

To analyze the spin texture, we decompose the composite boson
field $\phi ^{\alpha }$ into the two fields $\phi $ and
$n^{\alpha }$,
\begin{equation}
\phi ^{\alpha }(\bbox{x}) = \phi (\bbox{x})n^{\alpha }(\bbox{x}),
\qquad \phi (\bbox{x}) = e^{i\chi (\bbox{x})}\sqrt {\rho
(\bbox{x})}. \label{SpinA} 
\end{equation}
We substitute (\ref{SpinA}) into the density operator, obtaining
$\rho (\bbox{x}) = \phi ^{\dagger }(\bbox{x})\phi (\bbox{x})$ and
$\bbox{n}^{\dagger }(\bbox{x})\bbox{n}(\bbox{x})=1$, where
$\bbox{n}(\bbox{x})=(n^{\uparrow },n^{\downarrow })$. The spin
generator is expressed as $S^{a}(\bbox{x}) = \rho
(\bbox{x})\bbox{n}^{\dagger }(\bbox{x}){\tau
^{a}}\bbox{n}(\bbox{x})$. We count the number of the real fields
in the decomposition (\ref{SpinA}). The composite boson $\phi
^{\alpha }$ has four real fields in total, and the U(1) field
$\phi $ has two real fields. Hence, the two-component complex
field $n^{\alpha }$ has only two real fields. Such a field is the
SU(2) complex projective field and abbreviated to the CP$^1$
field. (In general the SU(N) complex projective field is
abbreviated to the CP$^{N-1}$ field.)

The ground state $|g_{0}\rangle $ satisfying (\ref{QHvacuu}) is
a coherent state of the CP$^1$ field,
\begin{equation}
\bbox{n}(\bbox{x})|g_{0}\rangle = \pmatrix{1\cr
0\cr}|g_{0}\rangle = {1\over \sqrt {2}}T\pmatrix{1\cr
1\cr}|g_{0}\rangle , \label{gZeroStateA} 
\end{equation}
where $T$ transforms the two-component electron field 
$(\psi _{1},\psi _{2})$ into $(\psi^{\uparrow},\psi^{\downarrow})$ 
as in (\ref{LayerToSAS}),
\begin{equation}
T = {1\over \sqrt {2}}
\pmatrix{1 &\phantom{-}1\cr
         1 &-1\cr},
\qquad T^{\dagger }T= 1.
\label{OperaT}
\end{equation}
The basic properties are 
$T^{\dagger }\tau ^{1}T=\tau ^{3}$, 
$T^{\dagger }\tau ^{2}T=-\tau ^{2}$ and 
$T^{\dagger }\tau ^{3}T=\tau ^{1}$.  
A spin texture is given by performing an SU(2) transformation,
$|\Phi \rangle =e^{i{\cal O}}|g_{0}\rangle $. It is a coherent
state of the CP$^1$ field. We may parametrize it as
\begin{equation}
\bbox{n}(\bbox{x})|\Phi \rangle = {1\over \sqrt {2}}T
\pmatrix{e^{i\varphi (\bbox{x})/2}\sqrt {1+\sigma (\bbox{x})}\cr
e^{-i\varphi (\bbox{x})/2}\sqrt {1-\sigma (\bbox{x})}}|\Phi
\rangle . \label{CoherCP} 
\end{equation}
In terms of the sigma field we have
\begin{eqnarray} s^{1}(\bbox{x})&\equiv &{1\over \rho
_{0}}\langle \Phi |S^{1}(\bbox{x})|\Phi \rangle = \sigma
(\bbox{x}) \nonumber\\ s^{2}(\bbox{x})&\equiv &{1\over \rho
_{0}}\langle \Phi |S^{2}(\bbox{x})|\Phi \rangle = -\sqrt
{1-\sigma ^{2}(\bbox{x})}\sin\varphi (\bbox{x}) , \nonumber\\
s^{3}(\bbox{x})&\equiv &{1\over \rho _{0}}\langle \Phi
|S^{3}(\bbox{x})|\Phi \rangle = \sqrt {1-\sigma ^{2}(\bbox{x})}
\cos\varphi (\bbox{x}) . \label{ClassPS} 
\end{eqnarray}
It is a characteristic feature of the coherent state that both
the density ($\sigma $) and its conjugate phase ($\varphi $) have
the classical fields. A generic ground state $|g\rangle $ of the
unperturbed Hamiltonian is given by the choice of $\sigma
(\bbox{x})=\sigma _{0}$ and $\varphi (\bbox{x})=\varphi _{0}$: In
particular, the state $|g_{0}\rangle $ is by $\sigma _{0}=\varphi
_{0}=0$.

Using the SU(2) transformation (\ref{SpinGenerU}) explicitly we
may express $s^{a}$ in terms of $f^{a}$,
\begin{equation} s^{a}(\bbox{x}) = {1\over \rho _{0}}\langle \Phi
|S^{a}(\bbox{x})|\Phi \rangle = s^{a}_{0} - \varepsilon
^{abc}f^{b}(\bbox{x})s^{c}_{0} + \cdots , \label{PseudClassX}
\end{equation}
with (\ref{QHvacuu}), where the dots $\cdots $ denote higher
order terms in $f^{a}$. Comparing this with (\ref{ClassPS}), we
can relate the functions $f^{a}$ to the fields $\sigma $ and
$\varphi $. We have $s^{1}(\bbox{x})=\sigma = -f^{2}$,
$s^{2}(\bbox{x})=-\varphi =f^{1}$ and $s^{3}(\bbox{x})=1$, up to
the first order in $f^{a}$. Hence, the perturbative expansion
around the mean-field ground state ($\sigma =\varphi =0$)
corresponds to the expansion in $f^{a}$.

The spin texture is classified topologically by the Pontryagin
number. The topological current, $Q_{\mu }={1\over 2\pi
}\varepsilon _{\mu \nu \lambda }\partial _{\nu }\bbox{n}^{\dagger
}\partial _{\lambda }\bbox{n}$, conserves trivially, $\partial
_{\mu }Q_{\mu }=0$. For the state (\ref{ClassPS}) it yields
\begin{equation}
\langle Q_{0}(\bbox{x})\rangle = {1\over 8\pi }\varepsilon
_{abc}\varepsilon _{ij}s_{a}\partial ^{i}s_{b}\partial ^{j}s_{c}
= {1\over 4\pi }\varepsilon _{ij}\partial _{i}\sigma \partial
_{j}\varphi . \label{PontrNumbe} 
\end{equation}
The topological charge $Q=\int d^{2}x Q_{0}(\bbox{x})$ is the
Pontryagin number, and topological excitations are Skyrmions
\cite{Skyrmion}. The classical Skyrmion minimizes the nonlinear
$\sigma $-model Hamiltonian (\ref{EnergChang}) in the
SU(2)-invariant limit. 

\section{LLL Projection and W$_{\infty
}\times $SU(2) Algebra}

When the magnetic field is strong enough, the magnetic energy
greatly exceeds thermal and potential energies. It is reasonable
to assume that electrons are confined within the lowest\ Landau\
level. To make a consistent theory it is necessary to make the
LLL projection by quenching the kinetic term \cite{refLLL}.

For this purpose we decompose the composite-boson coordinate
$\bbox{x}$ into the center-of-mass coordinate
$\check{\bbox{X}}\equiv (\check{X},\check{Y})$ and the relative
coordinate $\check{\bbox{R}}=(\check{P}_{y},-\check{P}_{x})$,
where $\bbox{x}=\check{\bbox{X}}+\check{\bbox{R}}$ and
$\check{P}_{k}$ is given by (\ref{MomenCS}). They satisfy
$[\check{X}, \check{Y}] =-i$, $[\check{P}_{x}, \check{P}_{y}] =
i$.
We use checked quantities for composite-boson variables. Two
independent sets of harmonic oscillators are defined,
$\check{a} \equiv {1\over \sqrt
{2}}(\check{P}_{x}+i\check{P}_{y})$ and $\check{b} \equiv {1\over
\sqrt {2}}(\check{X} - i\check{Y})$,
where $[\check{a}, \check{a}^{\dagger }] = [\check{b},
\check{b}^{\dagger }] = 1$.

The LLL projection is to quench the kinetic energy term in the
Hamiltonian. In the composite boson theory it is achieved by
imposing the LLL condition,
\begin{equation}
\check{a}\phi _{\alpha }(\bbox{x})|\widehat{\Phi }\rangle
={1\over \sqrt {2}} (\check{P}_{x}+i\check{P}_{y})\phi _{\alpha
}(\bbox{x})|\widehat{\Phi }\rangle =0 , \label{LLLcondi}
\end{equation}
on the state $|\widehat{\Phi }\rangle $ in our Hilbert space.

We are concerned about the spin texture $|\Phi \rangle
=e^{i{\cal O}}|g_{0}\rangle $, where ${\cal O}$ is the SU(2)
generator (\ref{SpinGenerU}). We examine if this state belongs to
the lowest\ Landau\ level. For this purpose we examine if ${\cal
O}|\widehat{\Phi }\rangle $ belongs to the lowest\ Landau\ level\
when $|\widehat{\Phi }\rangle $ does. Since we have
\begin{equation} a(\bbox{x})\phi _{\alpha }(\bbox{x}){\cal
O}|\widehat{\Phi }\rangle = {1\over 2}\sum _{b}(\tau
_{b})_{\alpha \beta }a(\bbox{x})f^{b}(\bbox{x})\phi _{\beta
}(\bbox{x})|\widehat{\Phi }\rangle \not= 0, \label{LLLcondiSpin}
\end{equation}
the state ${\cal O}|\widehat{\Phi }\rangle $ does not belong to
the lowest\ Landau\ level. We make the LLL projection of the
operator ${\cal O}$ and the c-number functions $f^{a}(\bbox{x})$
as follows.

We do this in a systematic way. We first make a Fourier
transformation of $f^{a}(\bbox{x})$, $f^{a}(\bbox{x})={1\over
2\pi }\int d^{2}q f^{a}(\bbox{q})e^{i\bbox{x}\bbox{q}}$. The
problem of the LLL projection is reduced to that of the plane
wave $e^{i\bbox{x}\bbox{q}}$. We make normal ordering with
respect to $\check{a}$ and $\check{a}^{\dagger }$ as
\begin{equation} e^{i\bbox{q}\bbox{x}} = \exp[{1\over \sqrt
{2}}q\check{a}^{\dagger }] \exp[-{1\over \sqrt
{2}}q^{*}\check{a}] \langle\!\!\langle
e^{i\bbox{q}\bbox{x}}\rangle\!\!\rangle \bbox{ }, 
\end{equation}
where
\begin{equation}
\langle\!\!\langle e^{i\bbox{q}\bbox{x}} \rangle\!\!\rangle
\bbox{ }\equiv
e^{-(1/4)\bbox{q}^{2}}e^{i\bbox{q}\check{\bbox{X}}} ,
\end{equation}
with $q=q_{x}+iq_{y}$. The LLL projection is to quench the
operators $\check{a}$ and $\check{a}^{\dagger }$. Hence,
\begin{equation}
\widehat{f}^{a}(\bbox{x})=\int {d^{2}q\over(2\pi)^{2}}
f^{a}(\bbox{q}) \langle\!\!\langle
e^{i\bbox{q}\bbox{x}}\rangle\!\!\rangle \bbox{ }, 
\end{equation}
and
\begin{equation}
\widehat{{\cal O}} = \int d^{2}x
\widehat{f}^{a}(\bbox{x})S^{a}(\bbox{x})=\int d^{2}q
f^{a}(-\bbox{q})\widehat{S}^{a}_{\bbox{q}} , \label{ProjeGener}
\end{equation}
where
\begin{equation}
\widehat{S}^{a}_{\bbox{q}} \equiv \int {d^{2}x\over2\pi}
S^{a}(\bbox{x}) \langle\!\!\langle
e^{-i\bbox{q}\bbox{x}}\rangle\!\!\rangle \bbox{ } 
\end{equation}
is the LLL projected spin operator. The LLL projection of the
density operator $\rho $ is similarly defined, which we denote by
$\widehat{\rho }_{\bbox{q}}$ in the momentum space.

After the LLL projection the Hamiltonian contains no kinetic
energy term. Yet, the dynamics arises since the components of the
center-of-mass coordinate do not commute, $[\check{X},
\check{Y}]=-i$. Using this commutation relation it is easy to
verify that the projected operators $\widehat{\rho }_{\bbox{q}}$
and $\widehat{S}^{a}_{\bbox{q}}$ satisfy the W$_{\infty }\times
$SU(2) algebra,
\begin{eqnarray} &&[\widehat{\rho }_{\bbox{p}},\widehat{\rho
}_{\bbox{q}}]={i\over \pi }\widehat{\rho
}_{\bbox{p}+\bbox{q}}\sin\bigl[{\bbox{p}\!\wedge\!\bbox{q}\over
2}\bigr] e^{(1/2)\bbox{p}\bbox{q}}, \nonumber\\
&&[\widehat{S}^{a}_{\bbox{p}},\widehat{\rho }_{\bbox{q}}]={i\over
\pi }
\widehat{S}^{a}_{\bbox{p}+\bbox{q}}\bigl[{\bbox{p}\!\wedge\!\bbox{q}\over
2}\bigr] e^{(1/2)\bbox{p}\bbox{q}},\nonumber\\
&&[\widehat{S}^{a}_{\bbox{p}},
\widehat{S}^{b}_{\bbox{q}}]={i\over 2\pi }\varepsilon
^{abc}\widehat{S}^{c}_{\bbox{p}+\bbox{q}}
\cos\bigl[{\bbox{p}\!\wedge\!\bbox{q}\over
2}\bigr] e^{(1/2)\bbox{p}\bbox{q}} +{i\over 4\pi }\delta
^{ab}\widehat{\rho
}_{\bbox{p}+\bbox{q}}\sin\bigl[{\bbox{p}\!\wedge\!\bbox{q}\over
2}\bigr] e^{(1/2)\bbox{p}\bbox{q}}, \label{SUCommuC}
\end{eqnarray}
with $\bbox{p}\!\wedge\!\bbox{q}=\varepsilon _{ij}p_{i}q_{j}$.
This governs the dynamics. It is a generalization of the
W$_{\infty }$ algebra characterizing the QH system \cite{refLLL}.
We call it the density algebra for simplicity. It is important
that the spin operator $\widehat{S}^{a}$ and the density operator
$\widehat{\rho }$ do not commute: Their actions are related in a
complicated way. Because of this relation the spin rotation
affects the Coulomb term (\ref{SpinCoulo}) though it involves
only the total electron density $\rho $. 

\section{Ground State
and Spin Texture}

We make the LLL projection of various terms in the Hamiltonian,
\begin{equation}
\widehat{H}_{C}^{+} = \pi \int
{d^{2}q}V_{+}(\bbox{q})\widehat{\rho }_{-\bbox{q}}\widehat{\rho
}_{\bbox{q}} , \qquad \widehat{H}_{C}^{-} = 4\pi \int
{d^{2}q}V_{-}(\bbox{q})\widehat{S}^{1}_{-\bbox{q}}
\widehat{S}^{1}_{\bbox{q}},
\qquad \widehat{H}_{Z} = -2\pi \lambda \widehat{S}^{3}_{0} ,
\end{equation}
where $V_{\pm }(\bbox{q})$ is the Fourier transformation of the
potential $V_{\pm }(\bbox{x})$.

The unperturbed Hamiltonian $\widehat{H}_{C}^{+}$ is minimized
by requiring
\begin{equation}
\widehat{\rho }_{\bbox{q}}|g\rangle = 4\pi \rho _{0}\delta
(\bbox{q})|g\rangle . \label{QHcondiBL} 
\end{equation}
We can impose this condition because the gapless mode is absent
in the total density fluctuation \cite{EIjos}, as leads to the
incompressibility of the QH system. This agrees with a result of
the representation theory \cite{Matsuo} that the W$_{\infty }$
algebra with no central extension has merely the trivial vacuum
sector.

We have already noticed that the ground state is a coherent
state of the CP$^1$ field. Hence, we impose
\begin{eqnarray}
\bar{s}^{1}_{\bbox{q}}&\equiv & {1\over \rho _{0}}\langle
g|\widehat{S}^{1}_{\bbox{q}}|g\rangle = 2\pi \sigma _{0}\delta
(\bbox{q}), \nonumber\\ \bar{s}^{2}_{\bbox{q}}&\equiv & {1\over
\rho _{0}}\langle g|\widehat{S}^{2}_{\bbox{q}}|g\rangle =-2\pi
\sqrt {1-\sigma _{0}^{2}}\sin\varphi _{0} \delta (\bbox{q}) ,
\nonumber\\ \bar{s}^{3}_{\bbox{q}}&\equiv & {1\over \rho
_{0}}\langle g|\widehat{S}^{3}_{\bbox{q}}|g\rangle = 2\pi \sqrt
{1-\sigma _{0}^{2}}\cos\varphi _{0} \delta (\bbox{q}).
\label{TypeIIcondi} 
\end{eqnarray}
All these states are degenerate with respect to the Hamiltonian
$H_{C}^{+}$. The degeneracy is removed by introducing the Zeeman
(tunneling) term, which is to choose $\sigma _{0}=\varphi
_{0}=0$, or
\begin{equation}
\langle g_{0}|\widehat{S}^{a}_{\bbox{q}}|g_{0}\rangle = 2\pi \rho
_{0}\delta ^{a3}\delta (\bbox{q}) . \label{gZeroState}
\end{equation}
This corresponds to (\ref{QHvacuu}) after taking its Fourier
transformation.

We consider the spin texture $|\widehat{\Phi }\rangle
=e^{i\widehat{{\cal O}}}|g_{0}\rangle $. We evaluate $\langle
\widehat{\Phi }|\widehat{S}_{\bbox{q}}|\widehat{\Phi }\rangle $
and denote its Fourier transformation by $\rho
_{0}\bar{s}^{a}(\bbox{x})$. Using (\ref{SUCommuC}),
(\ref{QHcondiBL}) and (\ref{TypeIIcondi}) we obtain
\begin{equation}
\bar{s}^{a}(\bbox{x}) = \delta ^{a1} - \varepsilon ^{1ab}
e^{\bbox{-}(1/2)\nabla ^{2}}f^{b}(\bbox{x}) + \cdots .
\label{PseudClassZ} 
\end{equation}
We find $\bar{s}^{a}(\bbox{x})=s^{a}(\bbox{x})$ with
(\ref{ClassPS}) for a sufficiently smooth configuration. It is
also easy to find
\begin{equation}
\bar{\rho }(\bbox{x}) = 2\rho _{0} + \nu Q_{0}(\bbox{x}) + \cdots
, \label{SkyrmDensi} 
\end{equation}
where $\bar{\rho }(\bbox{x})$ is the Fourier transformation of
$\langle \widehat{\Phi }|\widehat{\rho }_{\bbox{q}}|\widehat{\Phi
}\rangle $ and $Q_{0}(\bbox{x})$ is the Pontryagin number density
(\ref{PontrNumbe}). When $Q=\int d^{2}x Q_{0}(\bbox{x})\not=0$,
the spin texture describes Skyrmion excitations. A Skyrmion is a
topological excitation realized in the coherent mode. This
equation says that the Skyrmion carries electric charge $-e\nu
Q$. It has a fractional charge in general. 

\section{Effective
Hamiltonian}

Relevant correlation functions are those of density operators
$\widehat{\rho }$ and $\widehat{S}^{a}$. Two point functions such
as $\langle \widehat{\Phi }|\widehat{\rho
}_{\bbox{p}}\widehat{\rho }_{\bbox{q}}|\widehat{\Phi }\rangle $
cannot be evaluated by the algebraic relation alone. In so doing
we need to deal with $\langle
g_{0}|\{\widehat{S}^{a}_{\bbox{p}},
\widehat{S}^{b}_{\bbox{q}}\}|g_{0}\rangle
$. We use the formula,
\begin{equation}
\{\widehat{S}^{a}_{\bbox{p}},
\widehat{S}^{b}_{\bbox{q}}\}=-{1\over 2\pi }\varepsilon
^{abc}\widehat{S}^{c}_{\bbox{p}+\bbox{q}}
\sin\bigl[{\bbox{p}\!\wedge\!\bbox{q}\over
2}\bigr] e^{(1/2)\bbox{p}\bbox{q}} +{1\over 4\pi }\delta
^{ab}\widehat{\rho
}_{\bbox{p}+\bbox{q}}\cos\bigl[{\bbox{p}\!\wedge\!\bbox{q}\over
2}\bigr] e^{(1/2)\bbox{p}\bbox{q}} +
:\{\widehat{S}^{a}_{\bbox{p}}, \widehat{S}^{b}_{\bbox{q}}\}: ,
\end{equation}
where
\begin{equation}
\langle
g_{0}|:\!\widehat{S}^{a}_{\bbox{p}}
\widehat{S}^{b}_{\bbox{q}}\!:|g_{0}\rangle
= {1\over 4}\delta ^{a3}\delta ^{b3}\langle
g_{0}|:\!\widehat{\rho }_{\bbox{p}}\widehat{\rho
}_{\bbox{q}}\!:|g_{0}\rangle , 
\end{equation}
since $|g_{0}\rangle $ is a coherent state of the CP$^1$ field.
We next use
\begin{equation}
\{\widehat{\rho }_{\bbox{p}},\widehat{\rho }_{\bbox{q}}\}={1\over
\pi }\widehat{\rho
}_{\bbox{p}+\bbox{q}}\cos\bigl[{\bbox{p}\!\wedge\!\bbox{q}\over
2}\bigr] e^{(1/2)\bbox{p}\bbox{q}}+:\{\widehat{\rho
}_{\bbox{p}},\widehat{\rho }_{\bbox{q}}\}: . 
\end{equation}
Since $|g_{0}\rangle $ is an eigenstate of $\widehat{\rho
}_{\bbox{p}}$ as in (\ref{QHcondiBL}), this yields
\begin{equation}
\langle g_{0}|:\!\widehat{\rho }_{\bbox{p}}\widehat{\rho
}_{\bbox{q}}\!:|g_{0}\rangle = 4(2\pi )^{2}\rho _{0}^{2}\delta
(\bbox{p})\delta (\bbox{q}) - 2\rho _{0}\delta
(\bbox{p}+\bbox{q})e^{-(1/2)\bbox{p}^{2}}. 
\end{equation}
Therefore, we obtain
\begin{equation}
\langle
g_{0}|\!\{\widehat{S}^{a}_{\bbox{p}},
\widehat{S}^{b}_{\bbox{q}}\}\!|g_{0}\rangle
= \cases{ \rho _{0}\delta
(\bbox{p}+\bbox{q})\exp\bigl[-{\ell_{B}^{2}\over
2}\bbox{p}^{2}\bigr] &\hbox{for}\quad a=b=1,2 \cr 2(2\pi
)^{2}\rho _{0}^{2}\delta (\bbox{p})\delta (\bbox{q})
&\hbox{for}\quad a=b=3 \cr 0 & \hbox{otherwise} } 
\end{equation}
Various correlation functions are calculated by using these
formulas.

We are ready to evaluate the Hamiltonian on the spin texture
$|\widehat{\Phi }\rangle $ in a sufficiently smooth
configuration, $\Delta E\equiv \langle \widehat{\Phi
}|\widehat{H}_{C}^{+}+\widehat{H}_{C}^{-}+\widehat{H}_{Z}|\widehat{\Phi
}\rangle $. We define the effective Hamiltonian density ${\cal
H}_{\rm{eff}}$ by $\Delta E=\int d^{2}x {\cal H}_{\rm{eff}}$.
After a straightforward calculation we obtain
\begin{equation} {\cal H}_{\rm{eff}} = {1\over 2}\rho _{E}\sum
_{a=1}^{2}[\bbox{\nabla }s^{a}(\bbox{x})]^{2}+ {1\over 2}\rho
_{A}[\bbox{\nabla }s^{1}(\bbox{x})]^{2} + {e^{2}\rho
_{0}^{2}\over 2C}s^{1}(\bbox{x})s^{1}(\bbox{x}) - \lambda
s^{3}(\bbox{x}) , \label{EnergChang} 
\end{equation}
with $C$ the capacitance, and $\rho _{E}\equiv \rho _{s}^{+}-\rho
_{s}^{-}$, $\rho _{A}\equiv \rho _{s}^{+}+\rho _{s}^{-}$;
\begin{equation}
\rho _{s}^{\pm} = {\nu \over 16\pi }\int {d^{2}q\over(2\pi)^{2}}
V_{\pm}(\bbox{q})\bbox{q}^{2}e^{-(1/2)\bbox{q}^{2}} .
\label{PspinStiff} 
\end{equation}
We have $\rho _{E}=\rho _{A}$ in the monolayer system with spins. 
This gives
the LG theory of the coherent mode, where $\rho _{A}$ and $\rho
_{E}$ describe the spin stiffness. The result agrees with the one
\cite{MG} found in the SMA. It describes the nonlinear sigma
model in the SU(2)-invariant limit, where the Skyrmion solutions
are known explicitly \cite{Skyrmion}. Although their sizes are
infinitely large, they are made finite by the Zeeman term
(capacitance and tunneling terms). 

\section{Goldstone Mode}

We approximate the effective Hamiltonian (\ref{EnergChang}) as
\begin{equation} {\cal H}_{\rm{eff}} = {\rho _{E}\over 2}
(\bbox{\nabla }\varphi )^{2} + {\rho _{A}\over 2}(\bbox{\nabla
}\sigma )^{2} + {e^{2}\rho _{0}^{2}\over 2C}\sigma ^{2} + \lambda
\rho _{0}(\sigma ^{2}+\varphi ^{2}) \label{EffecHamilBL}
\end{equation}
for small fluctuations of $\sigma $ and $\varphi $. The fields
$\varphi (\bbox{x})$ and $\sigma (\bbox{x})$ are classical
fields. However, the commutation relation, $[\rho _{0}\sigma
(\bbox{x}), \varphi (\bbox{y})] = i\delta (\bbox{x}- \bbox{y})$,
follows naturally. To derive it we evaluate the equation of
motion,
\begin{equation} i{d\widehat{S}^{a}_{\bbox{k}}\over
dt}=[\widehat{S}^{a}_{\bbox{k}}, \widehat{H}] . 
\end{equation}
We then take its expectation value by the state $|\widehat{\Phi
}\rangle =e^{i\widehat{{\cal O}}}|g_{0}\rangle $. The resulting
set of equations agree precisely with the Heisenberg equations of
motion of the Hamiltonian (\ref{EffecHamilBL}) provided that the
above commutation relation is imposed.

We may diagonalize the Hamiltonian (\ref{EffecHamilBL}) by way
of the Bogoliubov transformation,
\begin{equation} H_{\rm{eff}} = \int d^{2}k E(\bbox{k})\alpha
_{\bbox{k}}^{\dagger }\alpha _{\bbox{k}} ,\quad E_{\bbox{k}}^{2}
= \Bigl({\rho _{E}\bbox{k}^{2}\over \rho _{0}} + \lambda
\Bigr)\Bigl({\rho _{A}\bbox{k}^{2}\over \rho _{0}} + {e^{2}\rho
_{0}\over C} + \lambda \Bigr). 
\end{equation}
When we switch off the Zeeman (tunneling) term by setting
$\lambda =0$, the dispersion relation $E_{\bbox{k}}$ describes a
gapless mode. This is the Goldstone mode associated with the
spontaneous polarization of the spins. When $\lambda \not=0$ it
becomes a gapful mode and called a pseudo-Goldstone mode.

The effective Hamiltonian (\ref{EnergChang}) is valid for any
values of $C$ and $\lambda $ as far as the coherent mode
persists. In the bilayer system there are some important
comments. When $\lambda =0$ the dispersion relation reads
$E_{\bbox{k}}\approx e\sqrt {\rho _{E}/C}|\bbox{k}|$ as
$\bbox{k}\rightarrow 0$ due to the capacitance effect. It has a
linear dispersion relation leading to a superfluid mode. Thus,
the capacitance term turns the Goldstone mode into a superfluid
mode. However, when the capacitance term becomes too large, the
Halperin $(m,m,m)$ phase breaks down with the loss of quantum
coherence \cite{EIjos}. It is taken over by another Halperin
phase \cite{HalperinC}, that is the $(m_{1},m_{2},n)$ phase with
$m_{1}m_{2}\not=n^{2}$. 

\section{Electric Currents:}

We analyze the electric current in external electric field. We
first consider the monolayer system. In the unprojected space the
current may be defined as follows. We make an infinitesimal local
phase transformation, $\psi \rightarrow e^{if(\bbox{x})}\psi $,
with the gauge field fixed. Since this is not a symmetry, the
Hamiltonian is modified by $\Delta H_{f}= (1/e)\int d^{2}x
J_{i}\partial _{i}f$. This defines the current $J_{i}$, which is
essentially the N\"other current. By making the LLL projection we
obtain the formula
\begin{equation} k_{i}\widehat{J}_{i}(\bbox{k}) = {ie\over \hbar
}{\delta \Delta \widehat{H}_{f}\over \delta
f_{-\bbox{k}}}\biggm|_{f\rightarrow 0} \label{NoethLLL}
\end{equation}
in the momentum space. The generator of the local phase
transformation is a smeared density operator. It is
$\widehat{O}_{f}=\int d^{2}q f(-\bbox{q})\widehat{\rho
}_{\bbox{q}}$ after the LLL; see (\ref{ProjeGener}). The
Hamiltonian transforms as
\begin{equation}
\widehat{H} \rightarrow
e^{-i\widehat{O}_f}\widehat{H}e^{i\widehat{O}_f} \equiv
\widehat{H} + \Delta \widehat{H}_{f} . \label{StepA}
\end{equation}
Since $\widehat{O}_{f}$ is just a c-number on $|g\rangle $ due to
the condition (\ref{QHcondiBL}) we obtain $\langle g|\Delta
\widehat{H}_{f}|g\rangle =0$ from (\ref{StepA}) and hence
$J_{i}=0$ from (\ref{NoethLLL}). Therefore, no currents are
induced by the Coulomb and Zeeman terms.

However, we cannot apply this argument to the external electric
field term $H_{E}$. For simplicity we assume a constant field
$E_{i}$, and we choose the potential $A_{0}=-x_{i}E_{i}$. The
term $H_{E}$ is given by
\begin{equation}
\widehat{H}_{E} = e\int d^{2}x \widehat{A}_{0}(\bbox{x})\rho
(\bbox{x}) = e\int d^{2}q A_{0}(\bbox{q})\widehat{\rho
}_{-\bbox{q}} , 
\end{equation}
after the LLL projection. The term is ill defined on the state
$|g\rangle $, $\widehat{H}_{E}|g\rangle =4\pi \rho
_{0}A_{0}(\bbox{q}=0)|g\rangle $ since $q_{j}A_{0}(\bbox{q})=2\pi
iE_{j}\delta (\bbox{q})$. Hence, there is no reason that $\langle
g|\Delta \widehat{H}_{f}|g\rangle =0$ for $\widehat{H}_{E}$ in
(\ref{StepA}).

Indeed, we can calculate explicitly $\Delta \widehat{H}_{f}$
for $\widehat{H}_{E}$, which reads
\begin{equation}
\Delta \widehat{H}_{f}=-i[\widehat{O}_{f}, \widehat{H}_{E}] =
-ie\int {d^{2}q}A_{0}(\bbox{q})\int
{d^{2}k}f_{-\bbox{k}}[\widehat{\rho }_{\bbox{k}}, \widehat{\rho
}_{-\bbox{q}}] . 
\end{equation}
Substituting this into the formula (\ref{NoethLLL}) and using the
density algebra (\ref{SUCommuC}) we obtain
\begin{equation}
\widehat{J}_{i}(\bbox{k})=e^{2}\varepsilon _{ij} E_{j}
\widehat{\rho }_{\bbox{k}} , \label{HallLLL}
\end{equation}
for a homogeneous electric field. This is an operator identity.
The charge-current conservation reads
\begin{equation} i{d\over dt}\widehat{J}_{0}(\bbox{k}) =
[\widehat{J}_{0}(\bbox{k}), \widehat{H}] =
-k_{i}\widehat{J}_{i}(\bbox{k}) , 
\end{equation}
where $\widehat{J}_{0}(\bbox{k})\equiv -e\widehat{\rho
}_{\bbox{k}}$ is the charge density. Eq.(\ref{HallLLL}) gives the
familiar Hall current on the ground state $|g\rangle $.
Evaluating it on the spin texture, $\langle \widehat{\Phi
}|\widehat{J}_{i}(\bbox{k})|\widehat{\Phi }\rangle $, and taking
its Fourier transformation, we obtain
\begin{equation}
\bar{J}_{i}(\bbox{x}) = {e^{2}\nu \over 2\pi } \varepsilon
_{ij}E_{j}[1+Q_{0}(\bbox{x})] , \label{HallCurre} 
\end{equation}
where $Q_{0}(\bbox{x})$ is the Skyrmion density
(\ref{PontrNumbe}) and we have used (\ref{SkyrmDensi}). The
Skyrmion density is dependent of time in the presence of an
electric field. However, Skyrmions would actually be trapped by
impurities just as quasiparticles. A Skyrmion excitation is
observable as a localized static object by means of measuring the
Hall current distribution.

One might ask why we can use the state $|g\rangle $ in
evaluating the Hall current (\ref{HallLLL}) when $\langle
g|\widehat{H}_{E}|g\rangle $ is ill defined. Indeed, the ground
state is no longer given by $|g\rangle $ in the presence of a
homogeneous electric field. It is given by \cite{ShizuHall}
\begin{equation} |g_{D}\rangle = e^{iMO}|g\rangle ,\qquad O =
\varepsilon _{ij}E_{j}\int d^{2}x x_{i}\rho (\bbox{x}) ,
\label{Shizuya} 
\end{equation}
where $M$ is the electron mass which should be set zero
($M\rightarrow 0$) in the large Landau-level gap-energy limit.
Since $O$ generates an ordinary local U(1) transformation,
$|g_{D}\rangle $ contains states in various Landau levels: the
factor $e^{iMO}$ mixes Landau levels. It produces a nontrivial
term in the Hamiltonian from the kinetic term $H_{K}$ containing
the factor $1/M$. Indeed, this factor is determined based on the
requirement that the anomalous term in $H_{E}$ is canceled by the
term so produced. We should mention that the Hall current is
obtainable \cite{ShizuHall} by a direct evaluation of $\langle
g_{D} |J_{i}(\bbox{x})|g_{D}\rangle $ with $J_{i}\approx
(e/M)\psi ^{\dagger }P_{i}\psi $ since it involves the factor
$1/M$. Namely, it is the state $|g_{D}\rangle $ that supports a
drift current. Therefore, we should use $|g_{D}\rangle $ also in
evaluating the Hall current (\ref{HallLLL}). However, this
formula does not contain the factor $1/M$. To evaluate such a
quantity there exists no distinction between two states
$|g\rangle $ and $|g_{D}\rangle $ in the limit $M\rightarrow 0$.
This justifies the use of the state $|g\rangle $. We remark that
our formula (\ref{NoethLLL}) is quite general which we can apply
to a heterogeneous electric field and also to any kind of
currents. We conclude that the current flows via a Landau-level
mixing though the mixing is infinitesimal as $M\rightarrow 0$. We
believe that this is the precise statement for the naive argument
given in Ref.\cite{refLLL}.

The current (\ref{HallCurre}) is all that we have in the
monolayer system. In the bilayer system we can apply a different
electric field $E_{j}^{\alpha }$ at each layer $\alpha (=1,2)$.
We introduce currents $\bbox{J}^{\pm }=\bbox{J}^{1}\pm
\bbox{J}^{2}$ and fields $\bbox{E}^{\pm }={1\over
2}(\bbox{E}^{1}\pm \bbox{E}^{2})$. The current $\bbox{J}^{+}$ is
the one associated with the local U(1) phase transformation,
while $\bbox{J}^{-}$ is associated with the local SU(2)
transformation (\ref{SpinRotat}) with $f^{1}=f^{3}=0$ and
$f^{2}=2f(\bbox{x})$. The Hall currents $\bar{J}^{E\pm }_{i}$ are
given by (\ref{HallCurre}) with $\bar{J}_{i}=\bar{J}^{E\pm }_{i}$
and $E_{j}=E^{\pm }_{j}$. Using the formula (\ref{NoethLLL}) and
the coherent-state condition (\ref{TypeIIcondi}), we derive the
supercurrent
\begin{equation}
\bar{J}_{i}^{C-}(\bbox{x})=2e \rho _{E} \partial _{i}\varphi
(\bbox{x}) , 
\end{equation}
from the Coulomb term and we derive the tunneling current
\begin{equation}
\bar{J}^{z}(\bbox{x}) \equiv 2e\lambda \rho _{0}s^{2}(\bbox{x}) =
-2e\lambda \rho _{0}\sqrt {1-\sigma (\bbox{x})^{2}} \sin\varphi
(\bbox{x}), 
\end{equation}
from the tunneling term between the two layers. These are all
that we have in the bilayer system. They are derived in an
analogous way. 

\section{Conclusion}

We have analyzed the monolayer QH system with spin degrees of
freedom and the bilayer QH system in the $(m,m,m)$ phase. Based
on the bosonic CS gauge theory with the LLL projection we have
presented a systematic way to investigate the quantum coherence
spontaneously developed in these systems. We have mapped the
bilayer system to the monolayer system. In particular, the
tunneling term is identified with the Zeeman term. The
capacitance term $H_{C}^{-}$ is specific to the bilayer system.
As the interlayer distance $d$ increases its existence becomes
crucial, and eventually the $(m,m,m)$ phase breaks down together
with the loss of quantum coherence. It is the capacitance term
and not the tunneling term that breaks the quantum coherence in
the bilayer system. This would explain the experimental fact
\cite{Sheena} that the quantum coherence has been observed at a
strong tunneling interaction such as $\Delta _{SAS}\approx 8.5$K.

When the Coulomb interaction term $H_{C}^{+}$ dominates the
system, these two systems have the SU(2) symmetry which is
spontaneously broken. The Zeeman term (the tunneling and
capacitance terms) selects the ground state $|g_{0}\rangle $ out
of many degenerate states $|g\rangle $ of the unperturbed system
$H_{C}^{+}$. It is possible to select another state $|g'\rangle $
as the ground state by making experimental arrangements upon
which the QH effect is observed. This could be done by tilting
the external magnetic field in the monolayer system and by
applying a bias voltage between the two layers in the bilayer
system. Indeed, in the presence of a bias voltage the capacitance
term $H_{\rm C}^{-}$ is not minimized by (\ref{QHvacuu}) but by a
certain nonzero value of the density 
difference $2\rho _{0}s^{1}=\rho _{1}-\rho _{2}$ between the two layers.  
It can be changed continuously by changing the bias voltage.  
This freedom exists only in the $(m,m,m)$ phase. Experimental
checks of these phenomena will be the easiest way to verify the
existence of the coherent mode in the QH system. We would like to
propose such experiments.

We have derived the effective Hamiltonian governing the
dynamics of the coherent mode in a bilayer quantum Hall system
using the W$_{\infty }\times $SU(2) algebra extensively. We have
already shown elsewhere \cite{EzaFerro} that such a Hamiltonian
leads to quantum coherent phenomena including the Josephson-like
effect. We have also shown that Skyrmion excitations are
observable by measuring the Hall current distribution. We believe
that our field-theoretical method is a powerful tool to analyze
various aspects of the QH effect. Detailed calculations of the
present work will be reported in a forthcoming
paper\cite{EzaIQC}.

\end{document}